\documentclass[a4paper,11pt]{article}
\pdfoutput=1 
\usepackage{jinstpub} 
\usepackage{caption} 
\usepackage{glossaries} 
\usepackage{graphicx} 
\PassOptionsToPackage{hyphens}{url}\usepackage{hyperref} 
\hypersetup{
    colorlinks,
    linkcolor={red!50!black},
    citecolor={blue!50!black},
    urlcolor={blue!80!black}
}
\usepackage{lineno} 
\modulolinenumbers[5]
\usepackage{lipsum} 
\usepackage{sidecap} 
\usepackage{subfig}
\usepackage{xcolor} 

\title{The AIDA-2020 TLU: a Flexible Trigger Logic Unit for Test Beam Facilities.}

\author[a,1]{P. Baesso,\note{Corresponding author.}}
\author[a]{D. Cussans,}
\author[a]{J. Goldstein}

\affiliation[a]{University of Bristol,\\Tyndall Avenue, BS8 1TL,\\United Kingdom}

\emailAdd{paolo.baesso@bristol.ac.uk}

\abstract{The AIDA-2020 \gls{tlu} has been designed to be a flexible and easily configurable unit to provide trigger and control signals to devices employed during test beams, integrating them with the beam telescope.
The most recent iteration of the \gls{tlu} (v1E) has been re-designed within the \mbox{AIDA-2020} project to integrate with hardware used in beam facilities.\\
Configuration and communication with the \gls{tlu} are performed over Ethernet. It can be employed as a stand-alone unit or be deployed as part of the EUDAQ2 data acquisition framework, which allows it to connect to a wide range of LHC readout systems.\\
The \gls{tlu} can operate with a sustained particle rate of 1~MHz and with instantaneous rates up to 20~MHz. In the current firmware iteration, the unit can time-stamp incoming signals with a resolution of 1.5~ns.\\
The hardware, firmware and software designs of the \gls{tlu} are freely accessible and benefit from constant inputs and upgrades from experienced users. \gls{tlu} units have already been deployed successfully in beam lines at CERN and DESY.}

\keywords{Trigger concepts and systems (hardware and software), Data acquisition circuits, Detector control systems (detector and experiment monitoring and slow-control systems, architecture, hardware, algorithms, databases)\\
\newline
\newline
\begin{SCfigure}[50][h]
  \centering
  \includegraphics[width=0.12\textwidth]{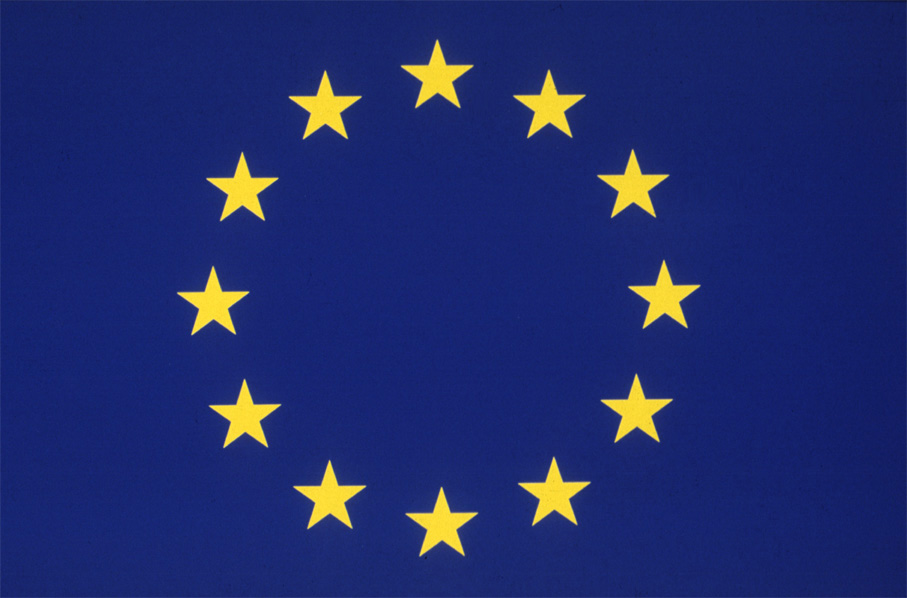}
    \caption*{This project has received funding from the European Union's Horizon 2020 research and innovation programme under grant agreement No 654168.}
\end{SCfigure}
}

\newacronym{ddye}{D$_{\text{dye}}$}{donor dye, ex. Alexa 488}
\newacronym[description={\glslink{r0}{F\"{o}rster distance}}]{R0}{$R_{0}$}{F\"{o}rster distance}
\newglossaryentry{r0}{name=\glslink{R0}{\ensuremath{R_{0}}},text=F\"{o}rster distance,description={F\"{o}rster distance, where 50\% ...}, sort=R}
\newglossaryentry{kdeac}{name=\glslink{R0}{\ensuremath{k_{DEAC}}},text=$k_{DEAC}$, description={is the rate of deactivation from ... and emission)}, sort=k}

\newacronym{af}{AF}{Auto Focus}
\newacronym{api}{API}{Application Programming Interface}
\newacronym{apt}{APT}{Advanced Positioning Technology}

\newacronym{ccd}{CCD}{Charge Coupled Device}
\newacronym{cdr}{CDR}{Clock and Data Recovery}

\newacronym{dac}{DAC}{Digital to Analog Converter}
\newacronym{daq}{DAQ}{Data Acquisition system}
\newacronym{dio}{DIO}{Digital InputOutput}
\newacronym{dpss}{DPSS}{Diode Pump Solid State}
\newacronym{dut}{DUT}{Device Under Test}

\newacronym{eeprom}{EEPROM}{Electrically Erasable Programmable Read-Only Memory}

\newacronym{fifo}{FIFO}{First In First Out}
\newacronym{fmc}{FMC}{FPGA Mezzanine Card}
\newacronym{fpga}{FPGA}{Field Programmable Gate Array}
\newacronym{fov}{FOV}{Field of View}

\newacronym{gpio}{GPIO}{General Purpose Input/Output}
\newacronym{gui}{GUI}{Graphic User Interface}

\newacronym{hdmi}{HDMI}{High-Definition Multimedia Interface}

\newacronym{i2c}{I$^2$C}{Inter-Integrated Circuit}
\newacronym{idc}{IDC}{Insulation-Displacement Contact}
\newacronym{ide}{IDE}{Integrated Development Environment}
\newacronym{ifd}{IFD}{Image File Directory}
\newacronym{ir}{IR}{Infra-red}

\newacronym{jtag}{JTAG}{Joint Test Action Group}

\newacronym{led}{LED}{Light Emitting Diode}
\newacronym{lhc}{LHC}{Large Hadron Collider}
\newacronym{lsb}{LSB}{Least Significant Bit}
\newacronym{lsm}{LSM}{Laser Sheet Microscopy}
\newacronym{lvds}{LVDS}{Low Voltage Differential Signaling}
\newacronym{lvpecl}{LVPECL}{Low-voltage Positive Emitter-Coupled Logic}

\newacronym{miso}{MISO}{Master In Slave Out}
\newacronym{mosfet}{MOSFET}{Metal Oxyde Semiconductor Field-Effect Transistor}
\newacronym{mosi}{MOSI}{Master Out Slave In}
\newacronym{msb}{MSB}{Most Significant Bit}

\newacronym{nan}{NaN}{Not a Number}
\newacronym{nas}{NAS}{Network Attached Storage}
\newacronym{ntc}{NTC}{Negative Temperature Coefficient}

\newacronym{pcb}{PCB}{Printed Circuit Board}
\newacronym{pll}{PLL}{Phase-Locked Loop}
\newacronym{pmt}{PMT}{Photo Multiplier Tube}
\newacronym{pwm}{PWM}{Pulse Width Modulation}
\newacronym{psf}{PSF}{Point Spread Function}

\newacronym{scl}{SCL}{Serial Clock Line}
\newacronym{sda}{SDA}{Serial DAta line}
\newacronym{sfp}{SFP}{Small Form-factor Pluggable}
\newacronym{shr}{SHR}{Synology Hybrid Raid}
\newacronym{sls}{SLS}{Selective Laser Sintering}
\newacronym{spi}{SPI}{Serial Peripheral Interface}
\newacronym{svn}{SVN}{Apache SubVersion}

\newacronym{tif}{TIFF}{Tagged Image File Format}
\newacronym{tlu}{TLU}{Trigger Logic Unit}
\newacronym{tps}{TPS}{Thinnest Part of the Sheet}

\newacronym{uart}{UART}{universal asynchronous receiver-transmitter}
\newacronym{uob}{UoB}{University of Bristol}
\newacronym{usb}{USB}{Universal Serial Bus}

\newacronym{rgb}{RGB}{Red Green Blue}
\newacronym{roi}{ROI}{Region Of Interest}

\begin{document}
\maketitle
\flushbottom



\section{Introduction}
The AIDA-2020~\cite{AIDA2020} project has the aim of advancing detector technologies by facilitating hardware development and testing across European institutes, universities and technology centres. Its program includes offering test beam and irradiation facilities where users can find a standardized selection of software tools, data acquisition systems and beam hardware.\\
Anyone who has been involved in beam tests of detectors will be familiar with the time involved in setting up their own hardware, integrating it with the beam line tools, defining the trigger and getting acquainted with the software before being able to take any data with the \gls{dut}. The AIDA-2020~\gls{tlu} is designed to minimise the setup time by providing a shared, flexible and programmable unit that can potentially substitute for tens of VME modules and a forest of cable connections.\\

\section{Overview}
\gls{dut}s can be connected to the \gls{tlu} by means of \gls{hdmi} ports. The \gls{tlu} can issue each \gls{dut} with trigger and synchronization signals as well as a system clock.\\
The unit accepts asynchronous input signals from external sources, such as beam scintillators, and can generate and distribute a global trigger based on any logical combination of the individual trigger inputs.
Upon issuing a global trigger to the \gls{dut}s, the unit stores the time-stamps of the individual inputs in an internal buffer to be read out over Ethernet.\\
\gls{dut}s can send signals back to the \gls{tlu} to indicate that they are busy. The response of the \gls{tlu} can be programmed, for instance stopping issuing any further trigger until the \gls{dut} de-asserts its busy line.\\
The \gls{tlu} has been designed to integrate with existing pixel beam telescopes that operate in EUDET/AIDA mode~\cite{RUBINSKIY2012923} and includes features that allow future expansion with other modes of operations.\\
Communications between the \gls{daq} and the \gls{tlu} use the IPBus~\cite{IpBusPaper} protocol, a well-established and reliable protocol widely used in the CMS~\cite{Chatrchyan:2008aa} and ATLAS~\cite{Aad:2008zzm} experiments at CERN. Amongst its main strengths, IPBus offers a reliability mechanism, which allows the client to detect and recover loss of packets.\\
A set of Python scripts is available to configure, control and read out the \gls{tlu}, allowing it to be used as a stand-alone tool. However, the full potential of the \gls{tlu} is achieved when used in conjunction with the EUDAQ2 data acquisition framework~\cite{EUDAQ2_2019}; EUDAQ2 is a powerful and modular framework that allows the users to integrate their own hardware with beam-telescopes software.
A dedicated software module (called the \emph{producer}) has been written for the \gls{tlu}, allowing the easy inclusion of the \gls{tlu} functionalities in any test facilities that provide the EUDAQ2 framework and hardware.

\section{Hardware}
\begin{figure}[t]
  \centering
  \includegraphics[width=1.0\textwidth]{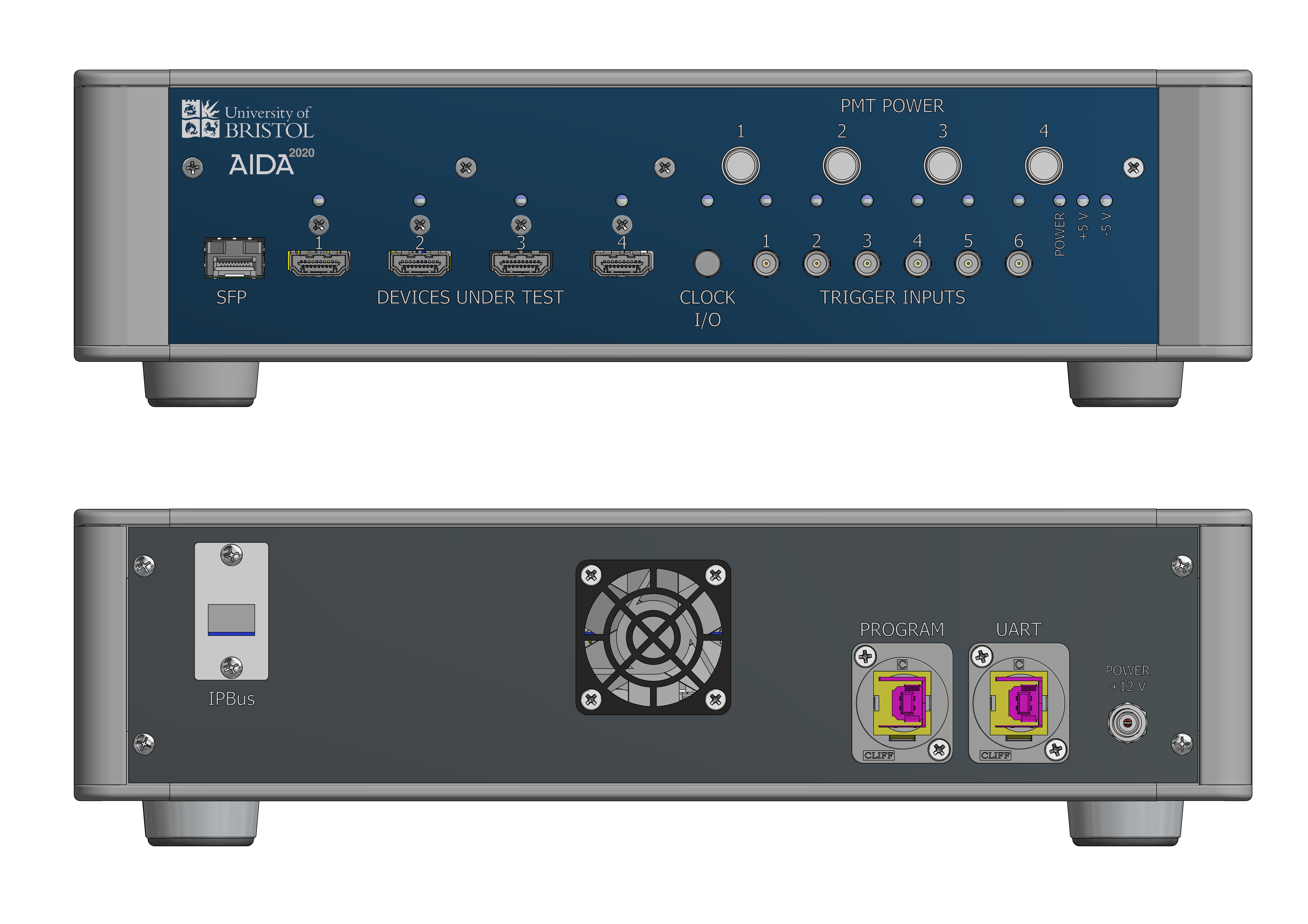}
  \caption{Sketch representation of the front (top) and back view (bottom) for the table-top enclosure. The unit requires 12~V to operate. The external dimensions are (W)~303~mm, (H)~75~mm, (D)~290~mm.}
  \label{fig:tableTopEnclosure}
\end{figure}

\begin{figure}[t]
  \centering
  \includegraphics[width=1.0\textwidth]{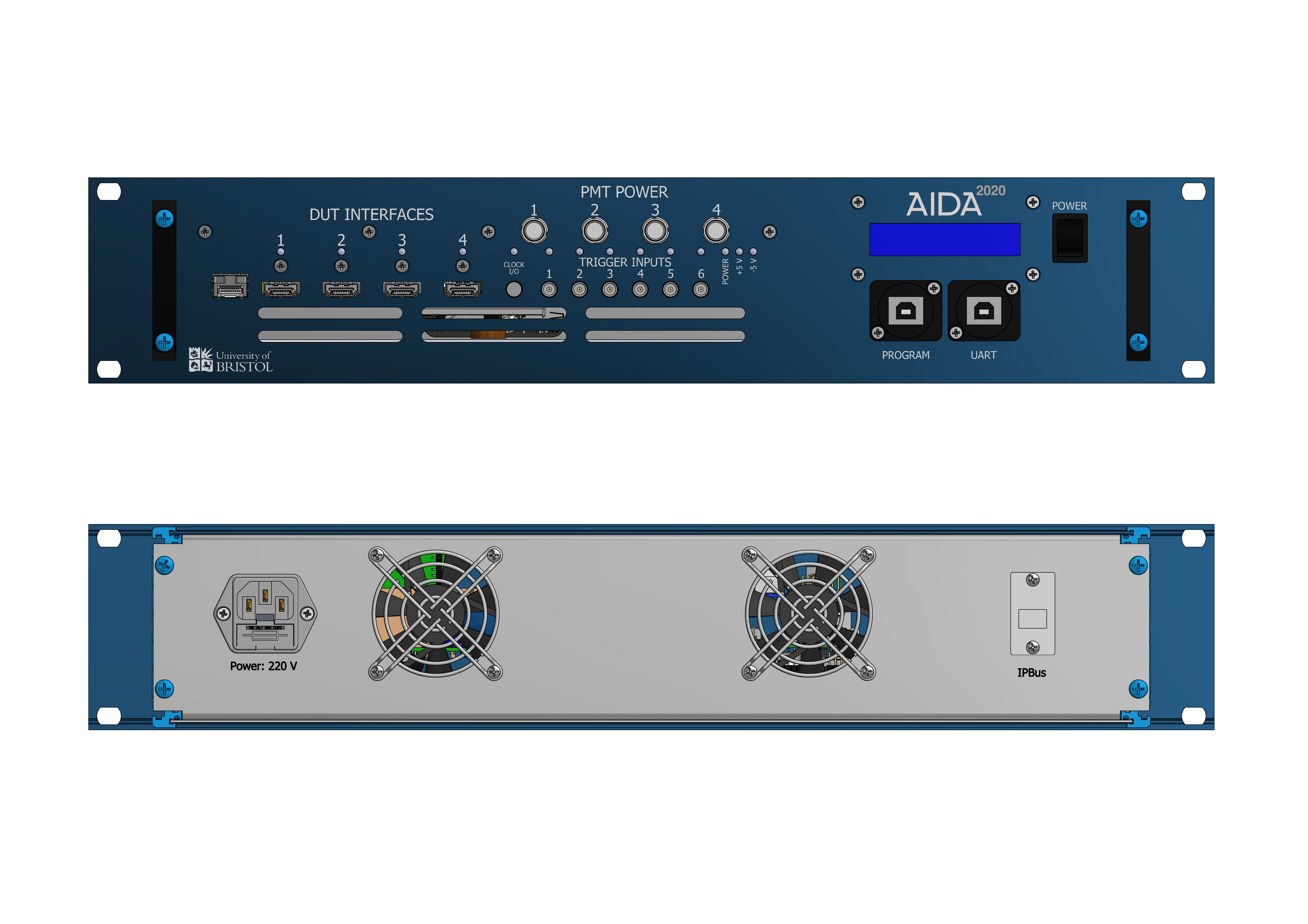}
  \caption{Sketch representation of the front (top) and back view (bottom) for the 19" rack mount. The hardware fits in a 2U unit and requires 220~V power. This enclosure includes a 19x2 characters display.}
  \label{fig:19InchEnclosure}
\end{figure}

The current version of the hardware (\emph{v1E}) is an evolution of the trigger logic unit initially designed to work with EUDET telescopes~\cite{CussansTLU,Jansen2016}. The core element of the unit is a \gls{pcb} mezzanine designed at the University of Bristol. The mezzanine has no programmable logic so it must be connected to a \gls{fpga} carrier board via a low pin count \gls{fmc} connector. The choice of having a separate carrier for the programmable logic was made to reduce the complexity of the design and maintain a high level of flexibility in terms of \gls{fpga} choice.\\
All the design and technical files are freely available on an open hardware repository~\cite{Ohwr_TLU}. A detailed user manual for the \gls{tlu} is available online~\cite{TLU_gitmanual}.\\
As part of the commitment to the AIDA-2020 project, a batch of units were manufactured and packaged for deployment in test beam facilities. These units include an Enclustra \gls{fpga} module based on the Xilinx Artix 7.\\
An auxiliary board, installed with all the \gls{tlu}s, can be used to provide power and control voltage to four photomultipliers.\\
Two different enclosures designs were produced for the \gls{tlu}: a table-top unit (figure~\ref{fig:tableTopEnclosure}) and a 19-inch rack mounted unit (figure~\ref{fig:19InchEnclosure}).
The table-top unit requires a 12~V supply to operate and, when fully configured, consumes about 15~W.
The rack-mounted unit is identical to the table-top one with the exception that it uses 220~V mains supply and has an additional display to visualize information.

\subsection{Interfaces}
The \gls{tlu} can be configured to generate trigger signals based on the combination of up to 6 inputs. The signals, usually provided by a \gls{pmt}, are routed to the \gls{tlu} using single-pole LEMO connectors\footnote{LEMO 00.250 family} and are fed to fast discriminators with independently configurable thresholds. The thresholds can be adjusted in the range [-1.3, +1.3]~V with 40~$\mu$V resolution by means of a 16-bit \gls{dac}. The input stage is protected by clamping diodes that limit the voltage within the range [-5, +5]~V.\\
Any of the 2$^{6}$ combinations of the inputs can be flagged as a valid word to issue a global trigger and multiple words can be defined in logical $OR$. This allows the user maximum flexibility in defining the trigger condition for their experiment.
One of the LEMO inputs can also be configured to receive a special synchronization signal (BEAM) from the beam-line. The user can choose to use the BEAM input as a special veto signal and program the \gls{tlu} to only issue triggers within a configurable time window relative to the reception of the BEAM, as described in section~\ref{ch:modes}.\\
The unit can provide 12~V power and control voltage to up to 4 photomultipliers by means of dedicated 9-mm LEMO connectors\footnote{LEMO part number EXP.0S.304.HLN.}. The control voltage can be configured in the range [0, +1]~V using 16-bit \gls{dac}s.\\
During normal operation the \gls{tlu} exchanges control signals with the \gls{dut}s according to one of the modes of operation described in Section~\ref{ch:modes}.
The \gls{dut}s are connected using four standard \gls{hdmi} connectors. All control signals (clock, trigger, busy, control, spare) are \gls{lvds} 3.3~V logic and use bi-directional half-duplex buffers so that any differential pair can be configured to feed signals from the \gls{dut} into the \gls{tlu}, or from the \gls{tlu} to the \gls{dut}. Additionally, the unit can provide +3.3~V power to the \gls{dut} on a dedicated pin of the connector.\\
To maintain retro-compatibility with existing hardware, the pin-out of the \gls{hdmi} connectors is the same as used in previous versions of the \gls{tlu} (table~\ref{tab:hdmi}).
\begin{table}[]
\footnotesize
\centering
\caption{HDMI pin mapping for the device under test. The pin-out is inherited from the CALICE ILC calorimeter project and has been kept unaltered to allow back-compatibility with the previous \gls{tlu}s.}
\label{tab:hdmi}
\begin{tabular}{|l|l|l|l|l|}
\hline
\textbf{HDMI PIN} & \textbf{TLU SIGNAL} & \textbf{} & \textbf{HDMI PIN} & \textbf{TLU SIGNAL} \\ \hline
1                 & $CLK$                 &           & 11                & $GND$                 \\ \hline
2                 & $GND$                 &           & 12                & $\overline{SPARE}$    \\ \hline
3                 & $\overline{CLK}$      &           & 13                & $n.c.$                \\ \hline
4                 & $CONTROL$             &           & 14                & $+3.3~V power$          \\ \hline
5                 & $GND$                 &           & 15                & $TRIGGER$             \\ \hline
6                 & $\overline{CONTROL}$  &           & 16                & $\overline{TRIGGER}$  \\ \hline
7                 & $BUSY$                &           & 17                & $GND$                 \\ \hline
8                 & $GND$                 &           & 18                & $n.c.$                \\ \hline
9                 & $\overline{BUSY}$     &           & 19                & $n.c.$                \\ \hline
10                & $SPARE$               &           &                   &                       \\ \hline
\end{tabular}
\end{table}\\
A \gls{sfp} interface can be used to connect transceivers to the \gls{tlu}. The data stream from the transceiver is processed by a \gls{cdr} and fed to the \gls{fpga}. The recovered clock can be used as a reference for the on-board clock chip. The \gls{sfp} interface has been used to successfully exchange data using a 250~MHz clock. This feature, not required in the AIDA-2020 mode of operation, allows the \gls{tlu} to be used as a timing distribution system, as described in Section~\ref{ch:dunemode}.\\
All communications between \gls{daq} and \gls{tlu} occur over a RJ45 interface directly connected to the \gls{fpga}: by using the IPBus protocol, the user can send configurations commands to the \gls{tlu} as well as retrieve status information and data from it.\\
In order to easily update the firmware, the unit has a \gls{usb} interface directly connected to the \gls{jtag} pins of the \gls{fpga}. A secondary \gls{usb} interface allows connection to the \gls{uart} port of the device to enable the possibility of connecting to the \gls{tlu} using a serial interface.\\
An additional two-pin LEMO connector\footnote{Part number EPG.00.302.NLN.} is configurable as either clock output or input, as explained in the next section.

\subsection{Clock capabilities}
The \gls{tlu} hosts an advanced clock multiplier and jitter attenuator~\cite{Si5345} that can be programmed to provide any clock frequency from 100~Hz to 1028~MHz. Users can modify the configuration file to generate any frequency suitable for their project.\\
Each \gls{dut} is connected to a dedicated clock output so that, potentially, each device can be operated with a different clock frequency. The clocks are distributed to the \gls{dut}s via the \gls{hdmi} connection. A clock signal is also available on a differential LEMO output on the front panel.\\
When using the \gls{tlu} as part of an AIDA-2020 test-beam infrastructure, the chip is configured by default to provide a 40~MHz clock. This clock is fed to the \gls{dut}s and used to provide time stamps for the trigger signals. The time stamps are stored in two separate registers, one containing a coarse stamp in 25~ns units and one containing a finer stamp in 1.56~ns units (obtained by clock-multiplication within the \gls{fpga}). The fine time-stamp is recorded separately for each input.\\
In order to lock the internal \gls{pll}, the \gls{tlu} requires a reference clock of 50~MHz; normally this is provided by an on-board crystal but in some scenarios it might be desirable to use a specific clock source to prevent random phase shifting. This can be achieved by changing the reference source: the unit can be configured to retrieve the reference from the differential LEMO on the front panel, from one of the \gls{dut} connectors, or from a clock retrieved from the \gls{sfp}.

\subsection{Firmware}
The current version of the firmware is designed to be deployed on an Enclustra Mars AX3 \gls{fpga} and uses an Artix 7 as target. The files are available from a Git repository~\cite{TLU_gw} so that any user can fork and modify them according to their needs.\\
The firmware contains a series of registers that can be addressed to configure any operational aspect of the \gls{tlu}.
Two software components of IPBus, the ControlHub and $\mu$HAL, allow the implementation of a reliability mechanism as well as providing an end-user \gls{api} to read and write the IPBus registers.\\

\subsection{Software}
In order to simplify the usage of the unit, the \gls{tlu} can be used in conjunction with the EUDAQ2 framework. To this purpose a \gls{tlu} producer is available in the EUDAQ2 Git repository~\cite{EUDAQ_git}.
As with any other EUDAQ2 producer, the hardware configuration can be defined by means of two configuration files parsed by the run control state machine. Details on the configuration parameters and their meaning are provided in the \gls{tlu} user manual.\\
The producer can access all \gls{tlu} registers via the abstraction provided by $\mu$HAL. It handles the commands from run control and provides high level configuration to the unit. It also reads the data buffer in the \gls{tlu} and sends the corresponding stream of data to the EUDAQ data formatter.\\
All the functionalities of the \gls{tlu} EUDAQ2 producer are also available in a Python command line interface. This allows users to test and debug the \gls{tlu} as a stand-alone unit, without having to install the EUDAQ2 framework.

\section{Modes of operation}\label{ch:modes}
\begin{figure}[h]
  \centering
  \includegraphics[width=0.650\textwidth]{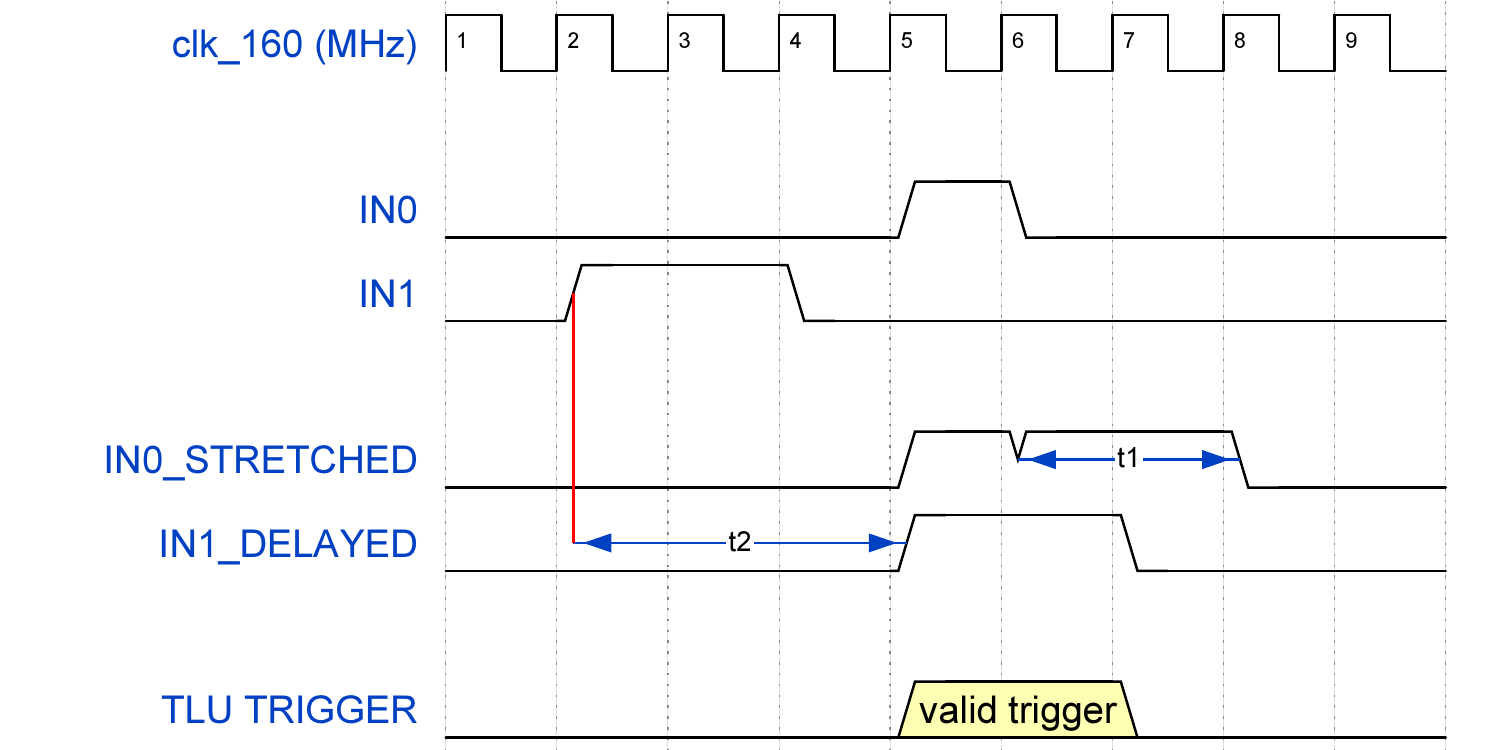}
  \caption{Example of trigger logic function: in this case only two inputs are considered. IN0 is stretched to increase its length by $t1$. IN1 is delayed by $t2$ to compensate for the different time of arrival. In this example it is assumed the trigger logic is a simple $AND$ of the two inputs.}
  \label{fig:triggerlogic}
\end{figure}

\begin{figure}[]
  \centering
  \includegraphics[width=1.0\textwidth, height=0.13\paperheight]{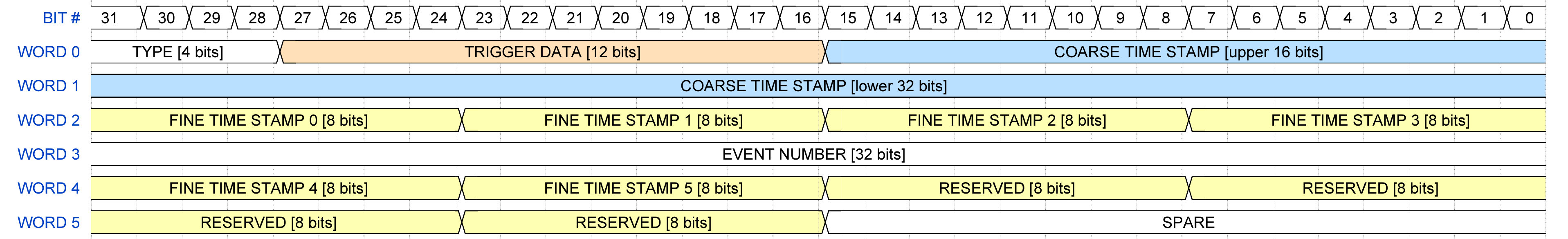}
  \caption{Event structure. Each trigger event is comprised of six 32-bit words. The unused words are intended for future firmware updates.}
  \label{fig:eventStructure}
\end{figure}

The \gls{tlu} offers a flexible and simple way of implementing any logical combination of the six input signals and producing a corresponding trigger output that is then distributed to the \gls{dut}.\\
Any of the $2^{6}= 64$ combinations of the input signals can be selected to provide a valid trigger function. To account for differences in cable lengths, the input signals can be stretched and/or delayed in 6.25~ns units (up to 190~ns) (figure~\ref{fig:triggerlogic}). Additionally, the unit can be configured to periodically generate internal triggers as a way to debug hardware without having to rely on external detectors.
When a valid trigger is formed, the \gls{tlu} creates an event data structure and stores it in an internal event buffer that can be read out via IPBus. Up to 1356 events can be stored in the buffer.
The structure of a data event is shown in figure~\ref{fig:eventStructure} and is based on six 32-bit words:
\begin{itemize}
  \item \emph{type} (4-bits) contains information about the type of trigger (for instance to distinguish internally generated triggers from triggers due to input signals).
  \item \emph{trigger data} (12-bits) indicates which of the six inputs were active when the trigger was generated.
  \item \emph{coarse time-stamp} (48-bits) is used to tag each trigger with respect to the system clock. Under normal conditions this counter increases in 25~ns units.
  \item \emph{fine time-stamp} (8-bits): for each of the inputs, a 5-bit fine time-stamp is stored and padded to 8-bits. This counter increases in 1.5~ns units.
  \item \emph{reserved} (8-bits) these bytes are reserved for future expansions of the \gls{tlu} hardware.
  \item \emph{spare} (32-bits) this word is reserved for future expansions of the EUDAQ2 producer.
\end{itemize}
During normal operation with EUDAQ2, the \gls{tlu} can sustain a continuous trigger rate of 1~MHz with instantaneous rates up to 20~MHz.\\

\subsection{Handshake}
\begin{figure}[]
  \centering
  \includegraphics[width=1.0\textwidth]{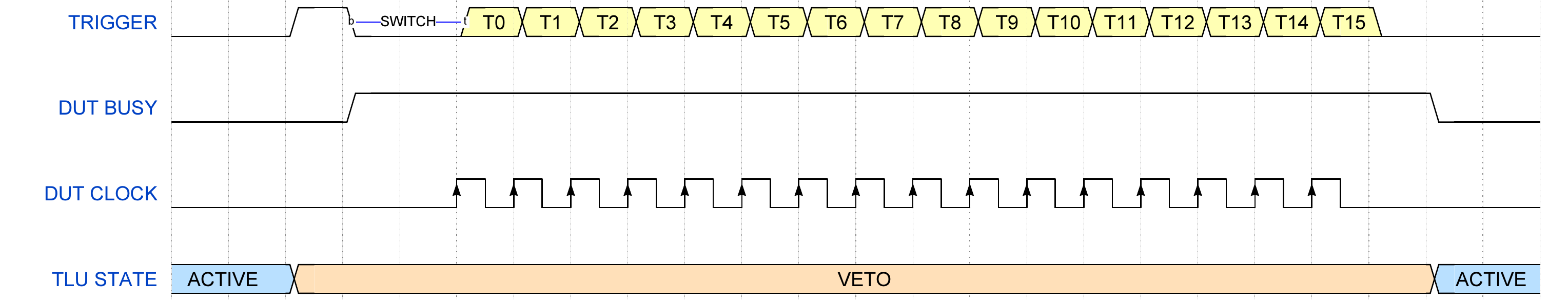}
  \caption{EUDET handshake mode: the \gls{tlu} is unable to issue any trigger until the handshake protocol is terminated. The total handshake has a duration of $\sim$450~ns}
  \label{fig:eudet}
\end{figure}

\begin{figure}[]
  \centering
  \includegraphics[width=1.0\textwidth]{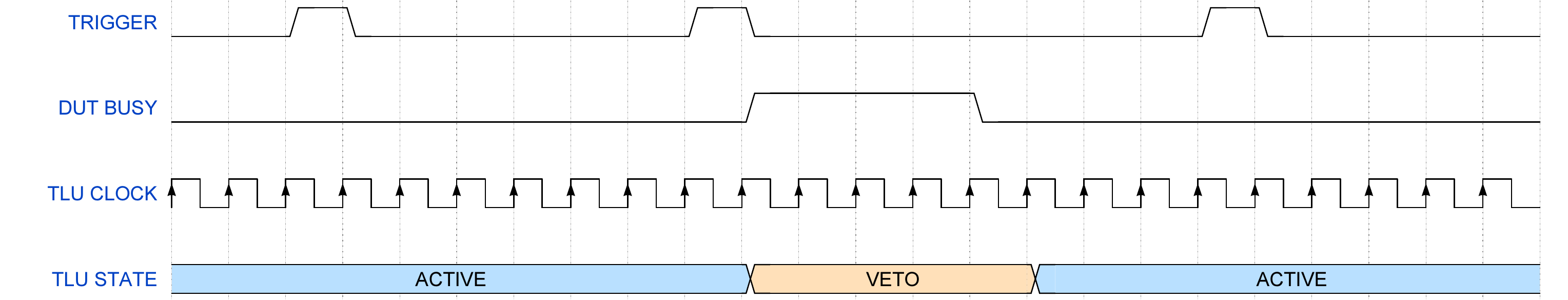}
  \caption{AIDA handshake mode. No data is exchanged between \gls{dut} and \gls{tlu}; upon the reception of a trigger, the unit is ready to issue new triggers after 1 clock cycle unless a device asserts its busy line.}
  \label{fig:aida}
\end{figure}

Valid trigger signals are propagated to any \gls{dut} connected and enabled. The \gls{tlu} can implement different handshake protocols depending on the type of device connected and the selected mode of operation. Two main modes of operation are traditionally referred to as EUDET mode and AIDA mode. EUDET mode is also referred to as \emph{trigger handshake mode}, while the AIDA mode is referred to as \emph{simple handshake mode}. Each \gls{dut} connected to the unit can operate in either mode independently.\\
When in EUDET mode (figure~\ref{fig:eudet}):
\begin{enumerate}
  \item Upon generating a valid trigger, the \gls{tlu} asserts the TRIGGER line.
  \item On receipt of the TRIGGER signal, the \gls{dut} asserts the BUSY line.
  \item The \gls{tlu} acknowledges the \gls{dut} being busy by de-asserting the TRIGGER line and switching its control to the output of a shift register containing the trigger number. As long as BUSY is active the \gls{tlu} cannot issue any further trigger to any device.
  \item The \gls{dut} controls the data transfer by toggling the DUT CLOCK line. Data changes on the rising edge and the least significant bit of the trigger data is shifted out first.
    Only the least significant 15 bits of the 32-bit trigger counter are clocked out. If more than 15 clock pulses are issued on the DUT CLOCK line the TRIGGER output is set to zero. The \gls{dut} should issue 16 clock pulses in order to clock out the 15 bits of the trigger number and return the TRIGGER line to logical low. This avoids glitches on the TRIGGER line when the \gls{dut} de-asserts the BUSY line.
  \item The \gls{dut} de-asserts the BUSY line whenever the shifting of the trigger number is completed and the detector is ready to accept a new trigger. At this point the system returns in the initial state.
\end{enumerate}
In EUDET mode the fastest trigger rate is limited by the slowest detector in the system. As long as one of the \gls{dut}s is busy, the \gls{tlu} is in a veto state and no further triggers can be issued.\\
However, it is possible to configure the \gls{tlu} to ignore the busy line on a specific device. In this case the handshake protocol will proceed in the same way as described, but that the BUSY line will affect only the \gls{dut} generating it. Therefore, the \gls{tlu} will still issue triggers to any device which is not in a busy state. This modified handshake mode allows the system to operate at a higher trigger rate but has the drawback that different devices in a run can receive a different number of triggers, thus requiring the data to be realigned off-line.\\
The AIDA mode of operation (figure~\ref{fig:aida}) adopts a simpler protocol:
\begin{enumerate}
  \item When a valid trigger is formed, the \gls{tlu} asserts TRIGGER synchronously with its internal clock
  \item The line is maintained for one clock cycle and then de-asserted. At this point the system is ready for a new trigger.
  \item If at any time the \gls{tlu} receives a BUSY signal from a \gls{dut}, it will veto all incoming triggers while the signal is active.
\end{enumerate}
Operating in AIDA mode means that no information is passed between the \gls{tlu} and the \gls{dut}s  with respect to the current trigger number. As such, normally it is necessary that all the devices register all the triggers received to ensure that the data can be correctly aligned at the end of the run. However, as in the EUDET mode, it is possible to configure the \gls{tlu} to ignore the BUSY signal from specific \gls{dut}s. Choosing this mode of operation means that the only limiting factor to the trigger rate is the capability of the \gls{tlu} of issuing triggers but has the drawback
that a \gls{dut} will record a only subset of the total number of triggers issued. Re-alignment of the data must necessarily be performed off-line, for instance by comparing the time stamps of the recorded triggers.
\\
\subsection{Shutter signal}
\begin{figure}[]
  \centering
  \includegraphics[width=1.0\textwidth]{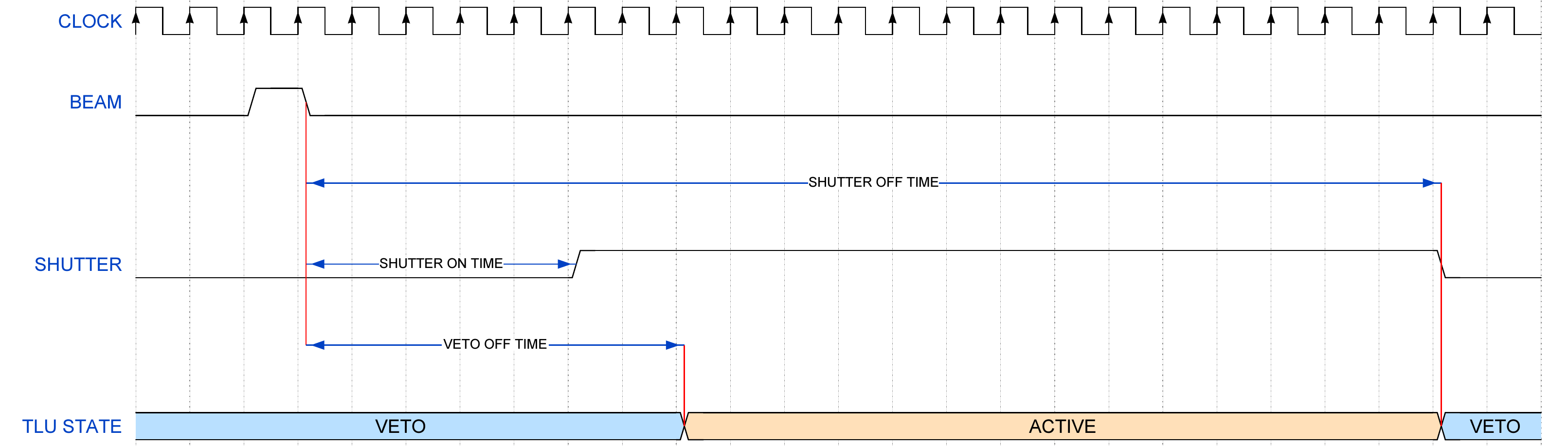}
  \caption{Diagram of operation of the timed-trigger mode (shutter mode). The BEAM signal is received by the \gls{tlu} and used as a time reference to open a virtual shutter. The shutter signal is issued to the \gls{dut} on the control line. The opening and closing of the shutter are determined by two programmable parameters (SHUTTER OFF and SHUTTER ON). While the shutter is open, an additional parameter (VETO OFF) defines the time window during which the \gls{tlu} can issue triggers to the \gls{dut}.}
  \label{fig:shutter}
\end{figure}

During standard operation, all inputs of the \gls{tlu} are interchangeable in terms of functionality. However, it is possible to assign special behaviour (beam) to one of the inputs in order to act as a reference to operate a virtual shutter within the unit (figure~\ref{fig:shutter}). The \gls{tlu} will issue the shutter signal to any \gls{dut} configured to receive it and only issue triggers while in the active part of the cycle. The \gls{dut} will start taking data only during a well-defined window with respect to the arrival of the beam signal. The feature can be useful if the device has limited buffer capability preventing it from taking data continuously.\\
The relative time position of the active shutter window with respect to the input signal is configured by two parameters (SHUTTER\_ON\_TIME and SHUTTER\_OFF\_TIME in the diagram). A third parameter (VETO\_OFF\_TIME) is used to define when, within the shutter active window, the \gls{tlu} can issue triggers to the \gls{dut}.

\section{Distribution and usage}
The version of hardware described in this work has already been distributed to CERN and DESY to be used in test beam campaigns.\\
A total of 19 units have been produced so far, 18 of which have been deployed in laboratories and beam lines. Following the request from several users, a new batch of 15-20 units will be manufactured in Q3-2019.\\
The first test of the AIDA-2020 \gls{tlu} in its intended role, i.e. providing timing and synchronization signals to a prototype detector using the beam-test infrastructure, was carried out during a TORCH~\cite{TORCH} hardware test in October-November 2017. During this campaign the \gls{tlu} was successfully used to synchronize data from the AIDA beam-telescope and a prototype from the TORCH detector R\&D project~\cite{HARNEW2018}. Since then, the AIDA-2020 TLU has been available at both DESY and CERN, typically in parallel with the old EUDET TLU to allow users to familiarize themselves with the new unit, while the AIDA TLU slowly replaces the old EUDET TLU in the beam line facilities.\\

\section{Use as a flexible master clock unit}\label{ch:dunemode}
\begin{figure}[]
  \centering
  \includegraphics[width=1.0\textwidth]{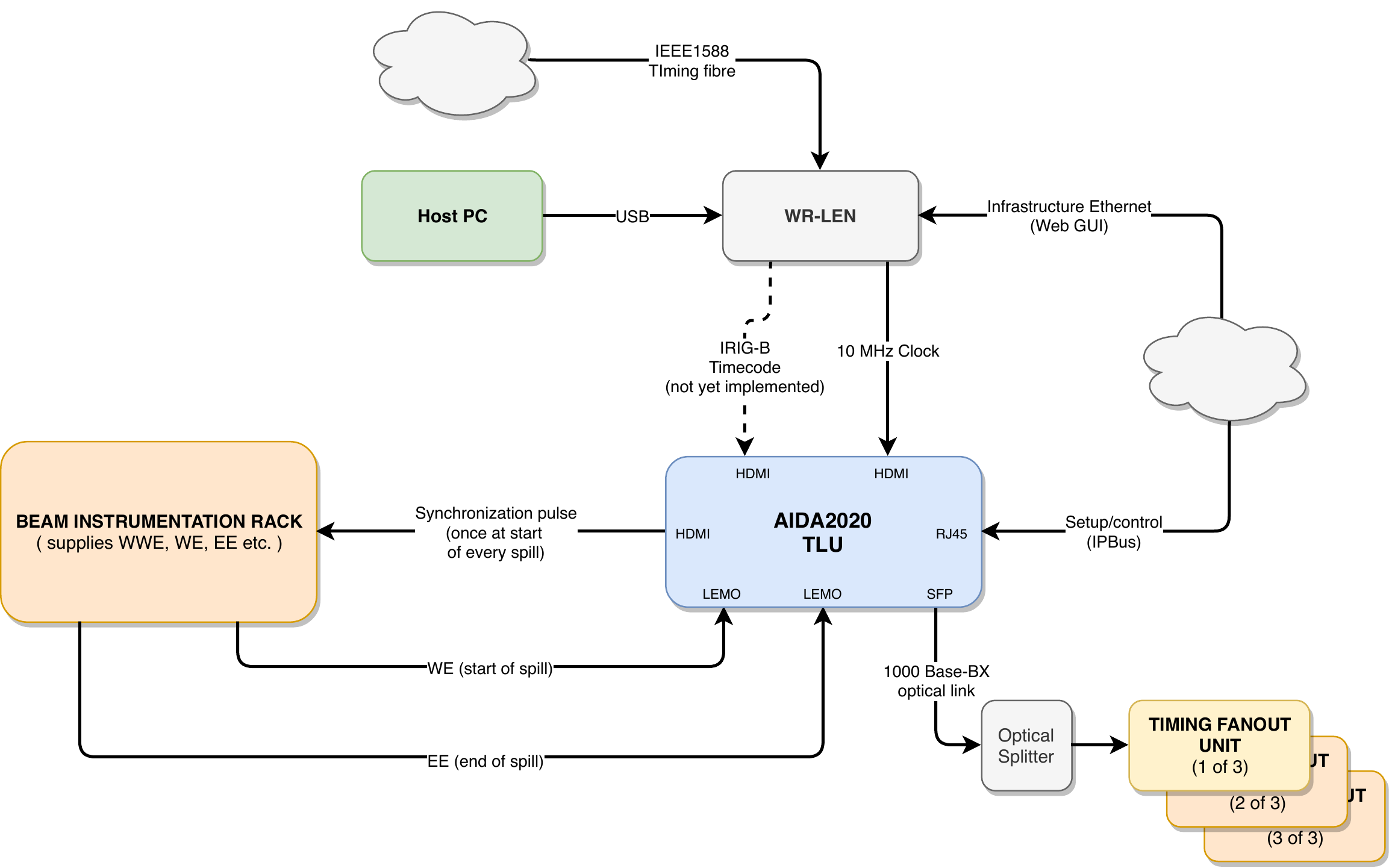}
  \caption{Schematic representation of the \gls{tlu} usage in the protoDUNE setup: the \gls{tlu} exchanges synchronization signals with the instrumentation rack, receives commands from the \gls{daq} and distributes an encoded data stream via its SFP interface.}
  \label{fig:protoDUNE}
\end{figure}

Thanks to the flexibility in its design, the \gls{tlu} was also adopted to act as master clock unit in the protoDUNE experimental setup~\cite{protoDUNE}. Figure~\ref{fig:protoDUNE} shows a schematic representation of the master synchronization distribution.\\
The unit receives synchronization and trigger commands from the software \gls{daq}, as well as a precise 10~MHz clock reference. The internal clock generator is used to produce a 250~MHz clock, phase-aligned with the 10~MHz one, which is then encoded with the data using 8b/10b encoding. The data stream is broadcast to the endpoints via a series of optical and electronic fan-out units.
For this application the unit is programmed with a dedicated firmware, developed in parallel to the one used for the AIDA2020 trigger unit.\\

\section{Conclusions}
The trigger logic unit developed for the AIDA-2020 program has successfully been deployed in real data-taking scenarios in several beam tests, allowing users to integrate their hardware with the EUDET telescopes. While it is fully back-compatible with the existing EUDET type telescopes, it introduces new modes of operation that increase the achievable trigger rate even in the presence of slow devices in the system.\\
The new unit is fully integrated in the EUDAQ2 framework, allowing users to easily set-up their hardware using a set of consistent tools. As EUDAQ2 is the chosen \gls{daq} at many beam-lines, including CERN and DESY, we expect that more and more users will soon become familiar with the AIDA-2020 TLU, abandoning the now out-of-date EUDET TLU.\\
The flexibility of the hardware, combined with its availability as an open-source project and the continuous feedback from users and beam-line operators, means that new features can be added to the unit making it a truly powerful and easy-to-use device in a beam-test.

\section{Acknowledgements}
The authors are particularly grateful to Jan Dreyling-Eschweiler and Yi Liu (DESY) for their help in developing and debugging the \gls{tlu} producer for EUDAQ2.\\
Andre Rummler (CERN) has also patiently tested the unit on the field and provided invaluable help to improve the software and to identify bugs.




\end{document}